\documentclass{article}
\usepackage{spconf,amsmath,graphicx,hyperref}

\usepackage{wrapfig}

\usepackage{cite}
\usepackage{amsmath,amssymb,amsfonts}
\usepackage{algorithmic}
\usepackage{graphicx}
\usepackage{textcomp}
\usepackage{xcolor}
\def\BibTeX{{\rm B\kern-.05em{\sc i\kern-.025em b}\kern-.08em
    T\kern-.1667em\lower.7ex\hbox{E}\kern-.125emX}}

\usepackage{amsmath}
\usepackage{amssymb}
\usepackage{color}
\usepackage{xspace}

\usepackage{threeparttable}
\usepackage{multirow}
\usepackage{tabularx}
\usepackage{tablefootnote}
\usepackage{booktabs}
\usepackage{makecell}
\usepackage{bbding}

\usepackage{amssymb}
\usepackage{pifont}

\usepackage{microtype}
\usepackage{graphicx}
\usepackage[caption=false,font=footnotesize]{subfig}
\usepackage{booktabs}
\usepackage{hyperref}
\usepackage{enumitem}

\newcommand{\todo}[1]{\textcolor{black}{#1}}

\newcommand{\benchmark}{\textsc{SingVERSE}\xspace}

\newcommand{\voicefixer}{Voicefixer\xspace}
\newcommand{\tfgridnet}{TFGridNet\xspace}
\newcommand{\nsnet}{NSNet2\xspace}
\newcommand{\frcrn}{FRCRN\xspace}
\newcommand{\demucs}{Demucs\xspace}

\newcommand{\llase}{LLaSE-G1\xspace}
\newcommand{\sgmseplus}{SGMSE+\xspace}
\newcommand{\storm}{StoRM\xspace}
\newcommand{\anyenhance}{AnyEnhance\xspace}

\title{\benchmark: A Diverse, Real-World Benchmark for Singing Voice Enhancement}
\name{\begin{tabular}{c}
    Shaohan Jiang$^{1\star}$ \qquad Junan Zhang$^{1\star}$ \qquad Yunjia Zhang$^{1\star}$ \\
    Jing Yang$^{2}$ \qquad Fan Fan$^{2}$ \qquad Zhizheng Wu$^{1}$
    \end{tabular}\thanks{$^\star$Equal Contribution. Names are in alphabetical order.}}

\address{$^{1}$The Chinese University of Hong Kong, Shenzhen \\
      $^{2}$Central Media Technology Institute, Huawei}
\begin{document}
%
\maketitle
\begin{abstract}

This paper presents a benchmark for singing voice enhancement. The development of singing voice enhancement is limited by the lack of realistic evaluation data. To address this gap, this paper introduces \textbf{\benchmark}, the first real-world benchmark for singing voice enhancement, covering diverse acoustic scenarios and providing paired, studio-quality clean references. Leveraging \benchmark, we conduct a comprehensive evaluation of state-of-the-art models and uncover a consistent trade-off between perceptual quality and intelligibility. Finally, we show that training on in-domain singing data substantially improves enhancement performance without degrading speech capabilities, establishing a simple yet effective path forward. This work offers the community a foundational benchmark together with critical insights to guide future advances in this underexplored domain. \todo{Demopage: \url{https://singverse.github.io}}
\end{abstract}
\begin{keywords}
Singing voice enhancement, speech enhancement, benchmark dataset, real-world audio processing
\end{keywords}

\section{Introduction}
\label{sec:intro}

\begin{figure}
    \centering
    \includegraphics[width=0.9\linewidth]{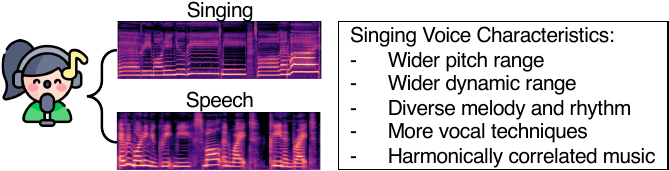}
    \caption{An illustration of the acoustic differences between singing (top) and speech (bottom). Singing is characterized by a wider pitch range and more stable harmonic structures corresponding to melody, which are distinct from the patterns in speech.}
    \label{fig:overview}
    \vspace{-0.6cm}
\end{figure}

The digital era has witnessed an explosion of recorded singing performances, from amateur home recordings to live concert captures, which are frequently captured in acoustically uncontrolled environments. Consequently, unlike professional studio productions, they often suffer from severe quality degradation due to ambient noise, complex reverberation, and interfering musical accompaniments. Enhancing these recordings is not only crucial for improving the listener's experience but also essential for enabling downstream applications such as audio remixing~\cite{fabbro2024sound} and automatic singing transcription~\cite{li2024robust}. While a common approach to this problem is to leverage state-of-the-art models from Speech Enhancement (SE)~\cite{zhang2025anyenhance,kang2025llase, wang2025metis}, singing voice exhibits a wider pitch and dynamic range, prolonged vowel phonations, and adherence to a specific rhythm and melody. Furthermore, the background in singing recordings is often harmonically correlated music, which poses a different challenge than the typically uncorrelated background noise in speech scenarios (as shown in Figure~\ref{fig:overview}). Therefore, applying models designed for speech directly to singing constitutes a significant domain mismatch, the impact of which has not been systematically investigated.

In this paper, our aim is to systematically investigate this underexplored area of singing voice enhancement. We begin by empirically demonstrating the aforementioned domain mismatch. Our controlled experiments, detailed in Section~\ref{sec:motivation}, confirm a significant performance distribution gap when applying standard SE models to singing versus speech, which validates the need for dedicated solutions and, crucially, for evaluation benchmarks that reflect real-world conditions.

However, existing evaluation practices on singing voice enhancement~\cite{xu2023mbtfnet, zhang2025anyenhance} fall short, as they largely rely on datasets synthesized from clean academic vocals with artificial noise and reverberation. Such data fails to capture real-world acoustic complexities. Consequently, a critical gap exists for a benchmark that is both \textbf{realistic} in its capture and \textbf{diverse} in its scenarios—essential for truly assessing a model's real-world performance. To fill this gap, we introduce \textbf{\benchmark} (\underline{Sing}ing-\underline{V}oice \underline{E}nhancement in \underline{R}eal \underline{S}c\underline{E}narios), a novel benchmark designed for realism and diversity. It consists of 7942 utterances across 18.36 hours, spanning 19 distinct acoustic scenarios, from reverberant concert halls to noisy roadsides, and provides corresponding studio-quality dry vocals as clean reference recordings for robust evaluation.

Leveraging \benchmark, we conduct a comprehensive evaluation of state-of-the-art enhancement models, uncovering crucial insights into their limitations. Our analysis first establishes that recording device quality is a fundamental factor influencing performance. Furthermore, it reveals a prevalent trade-off between perceptual quality and content intelligibility, which sorts models into distinct behavioral profiles (e.g., conservative or aggressive). Building on these insights, we demonstrate a straightforward yet effective path forward: simply training models on in-domain singing data significantly boosts their performance on singing enhancement without degrading their original speech enhancement capabilities. This work therefore provides the community with not only a foundational benchmark, but also crucial insights and a practical baseline to guide the development of the next generation of singing voice enhancement technologies.

Our main contributions are summarized as follows:
\begin{itemize}[itemsep=1pt,topsep=0pt,parsep=0pt]
    \item We introduce the first real-world benchmark \textbf{\benchmark} for singing voice enhancement. It covers diverse acoustic scenarios with various recording equipment, filling a critical gap for realistic evaluation.
    
    \item We perform a comprehensive evaluation of current state-of-the-art SE models and demonstrate an important way of using in-domain singing data for singing voice enhancement.
\end{itemize}

\section{Motivation: Quantifying the Domain Gap}
\label{sec:motivation}

Although speech enhancement (SE) has witnessed impressive progress, the vast majority of models are trained and validated on spoken corpora. Applying these models directly to singing voice constitutes a significant domain mismatch, but the performance degradation has not been systematically quantified. To address this, we first conduct a controlled experiment to empirically measure this gap. We construct two parallel test datasets that share identical acoustic degradations yet differ only in their source material: speech and singing.

\noindent\textbf{Experimental Design.}
To ensure a fair comparison, we select 255 English utterances each from the EARS dataset for \textit{speech}~\cite{richter2024ears} and the GTSinger dataset for \textit{singing}~\cite{zhang2024gtsinger}. We apply an identical degradation pipeline from \anyenhance~\cite{zhang2025anyenhance} to both sets, creating acoustically parallel corpora. We then evaluate nine state-of-the-art models: five discriminative (\voicefixer~\cite{liu2022voicefixer}, \tfgridnet~\cite{wang2023tf}, \nsnet~\cite{braun2020data}, \frcrn~\cite{zhao2022frcrn}, \demucs~\cite{defossez2019demucs}) and four generative (\llase~\cite{kang2025llase}, \sgmseplus~\cite{richter2023sgmse}, \storm~\cite{lemercier2023storm}, \anyenhance~\cite{zhang2025anyenhance}). Performance is quantified using three metrics: SpeechBERTScore (SBERT)~\cite{saeki2024speechbertscore} for semantics, speaker similarity (SIM) using a pretrained WavLM, and Word Error Rate (WER) from a Whisper-large-v3 model.

\begin{table}[htbp]
    \centering
    \small
    \renewcommand{\arraystretch}{0.9}
    \setlength{\tabcolsep}{3pt}
    \caption{Performance comparison of SE models on controlled speech vs. singing testsets. Results are shown as \textit{Speech} / \textit{Singing}. An asterisk (*) indicates a statistically significant ($p < 0.05$) degradation on the singing task. \todo{We evaluate two types of enhancement models: Discriminative (Disc.) and Generative (Gen.) models.}}
    \label{tab:motivation_results}
    \begin{threeparttable}
    \begin{tabular}{llccc}
        \toprule
        \textbf{Type} & \textbf{Model} & \textbf{SIM $\uparrow$} & \textbf{SBERT $\uparrow$} & \textbf{WER $\downarrow$} \\
        \midrule
        \multirow{5}{*}{Disc.} 
        & \voicefixer & 0.870 / 0.874 & 0.716 / 0.682\tnote{*} & 0.261 / 0.457\tnote{*} \\
        & \tfgridnet & 0.934 / 0.860\tnote{*} & 0.781 / 0.636\tnote{*} & 0.159 / 0.463\tnote{*} \\
        & \nsnet     & 0.934 / 0.886\tnote{*} & 0.738 / 0.656\tnote{*} & 0.188 / 0.486\tnote{*} \\
        & \frcrn      & 0.937 / 0.914\tnote{*} & 0.774 / 0.766 & 0.167 / 0.336\tnote{*} \\
        & \demucs     & 0.923 / 0.906\tnote{*} & 0.738 / 0.720\tnote{*} & 0.216 / 0.498\tnote{*} \\
        \midrule
        \multirow{4}{*}{Gen.} 
        & \llase   & 0.883 / 0.890 & 0.632 / 0.671 & 0.527 / 0.751\tnote{*} \\
        & \sgmseplus     & 0.928 / 0.910\tnote{*} & 0.759 / 0.741\tnote{*} & 0.249 / 0.430\tnote{*} \\
        & \storm      & 0.925 / 0.886\tnote{*} & 0.689 / 0.660\tnote{*} & 0.287 / 0.523\tnote{*} \\
        & \anyenhance & 0.961 / 0.949\tnote{*} & 0.835 / 0.830 & 0.142 / 0.300\tnote{*} \\
        \bottomrule
    \end{tabular}
    \end{threeparttable}
    \vspace{-10pt}
\end{table}

\noindent\textbf{Findings.}
The results in Table~\ref{tab:motivation_results} reveal a significant performance drop when models transition from speech to singing. This degradation is statistically significant ($p < 0.05$) in 22 of 27 comparisons, as confirmed by a Wilcoxon signed-rank test. This provides strong quantitative evidence for the domain mismatch, showing that models optimized for speech are ill-equipped for singing. However, the synthetic nature of this evaluation may not fully capture real-world complexities, highlighting the critical need for a benchmark with genuine, in-the-wild recordings to drive progress.

\section{\benchmark: A Real-World Benchmark}
\label{sec:benchmark}

The experiment in Section~\ref{sec:motivation} confirmed the performance gap but also highlighted the limitations of synthetic test data. To bridge this critical gap, we construct and release \textbf{\benchmark}, the first benchmark for singing voice enhancement designed to reflect real-world acoustic complexities. It features recordings from diverse, authentic environments and provides paired, studio-quality clean references for robust evaluation.

To achieve both diversity and realism, we gathered recordings from 18 distinct everyday environments (e.g., roadsides, KTVs) and concert venues. For the everyday recordings, ten students (5 male, 5 female) from music academies performed 30 songs, with each song captured in 3-4 different scenarios using a variety of professional (e.g., condenser microphones) and non-professional (e.g., smartphones, tablets) devices. For concert recordings, ``noisy'' tracks were sourced from public video platforms, while their corresponding clean, studio-grade vocal stems were extracted from official releases using an optimized SingNet pipeline~\cite{gu2025singnet}. All degraded clips are manually time-aligned with their clean references and segmented.

\begin{table}[htbp]
\centering
\small
\caption{Statistics of the \benchmark dataset. The values for Professional (Pro) and Non-Professional (Non-pro) devices are shown as Pro/Non-pro.}
\label{tab:dataset}
\renewcommand{\arraystretch}{0.9}
\setlength{\tabcolsep}{3pt}
\begin{tabular}{lccc}
\toprule
\multirow{2}{*}{\textbf{Scenario}} & \multicolumn{3}{c}{\textbf{Device: Pro / Non-Pro}} \\
\cmidrule{2-4} 
& \textbf{Utterances} & \textbf{Avg.(s)} & \textbf{Total(s)} \\
\midrule
Concert Hall & 62/94 & 9.8/8.8 & 611/832 \\
Concert & 112/335 & 10.1/11.8 & 1138/3965 \\
Arcade & 84/84 & 8.2/8.2 & 688/688 \\
Restaurant & 39/39 & 5.8/5.8 & 228/228 \\
Rehearsal Room & 120/120 & 8.2/8.2 & 439/439 \\
Classroom & 66/66 & 6.6/6.6 & 439/439 \\
Basement & 39/39 & 5.8/5.8 & 228/228 \\
Roadside & 384/368 & 8.2/8.2 & 3179/3036 \\
KTV & 230/180 & 7.6/7.7 & 1767/1394 \\
Piano Room & 178/178 & 8.2/8.2 & 1469/1469 \\
Passageway & 97/108 & 7.6/8.2 & 746/859 \\
Staircase & 64/65 & 7.4/7.4 & 476/481 \\
Meeting Room & 22/22 & 8.6/8.6 & 190/190 \\
Parking Lot & 68/68 & 8.9/8.9 & 606/606 \\
In the Car & 39/39 & 5.8/5.8 & 228/228 \\
Shopping Mall & 60/40 & 7.4/7.5 & 442/301 \\
Dormitory & 0/96 & 0/9.83 & 0/943.99 \\
Office & 60/60 & 7.3/7.3 & 442/442 \\
Park & 123/123 & 7.4/7.4 & 917/917 \\
\bottomrule
\end{tabular}
\vspace{-15pt}
\end{table}

Ultimately, \benchmark\ consists of \textbf{3,971 pairs} of real-world singing clips, totaling approximately \textbf{18.36 hours}. Its comprehensive coverage of acoustic scenarios is detailed in Table~\ref{tab:dataset}. This dataset serves as the foundation for the in-depth analysis in the following section.

\section{Analysis on \benchmark}
\label{sec:experiments}

Equipped with \benchmark, we move beyond synthetic tests to conduct a comprehensive analysis of state-of-the-art SE models in real-world conditions. Our investigation follows a three-step process: we first perform a benchmark-wide evaluation to uncover general performance patterns, then delve deeper to identify the primary factors driving performance, and finally, leverage these insights to propose and validate an effective enhancement strategy.

\noindent\textbf{Models.} We evaluate the same nine state-of-the-art systems introduced in our motivation analysis.

\noindent\textbf{Evaluation Metrics.} Our evaluation assesses two crucial aspects. For content preservation, we use SBERT, SIM, and WER. For perceptual quality, we employ two non-intrusive models: DNSMOS~\cite{reddy2022dnsmos} (SIG, BAK, OVRL) and NISQA~\cite{mittag2021nisqa}.

\noindent\textbf{Data Subsets.} To isolate the impact of recording quality, we partition \benchmark\ into Professional-Device (Pro.) and Non-Professional-Device (Non-pro.) subsets based on objective evaluations. We evaluate all models on both subsets, resulting in 9 models $\times$ 2 subsets $\times$ 19 scenarios = 342 distinct evaluations.

\subsection{Impact of Device Quality on Enhancement Results}
\label{sec:device_impact_analysis}

\begin{figure}[h]
    \centering
    \includegraphics[width=1\linewidth]{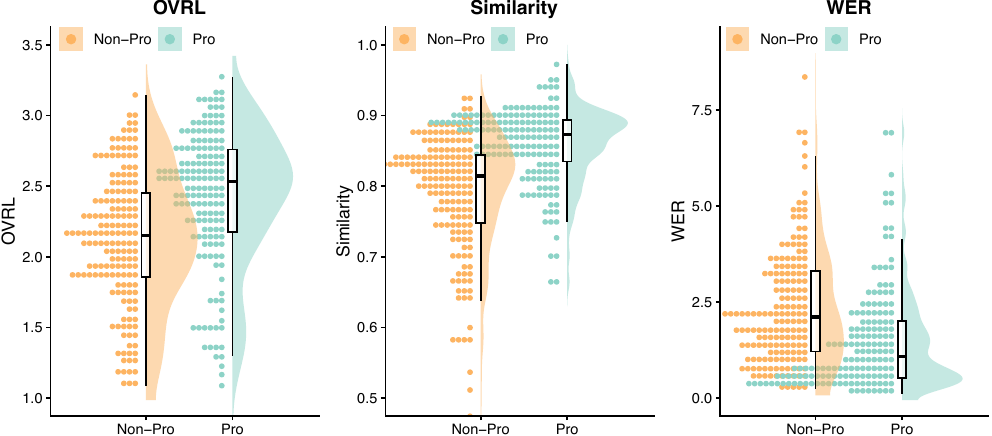}
    \caption{Performance distributions on Overall Quality (OVRL), Speaker Similarity, and Word Error Rate (WER) across Professional (Pro) and Non-Professional (Non-pro) device subsets. Higher is better for OVRL and Speaker Similarity; lower is better for WER.}
    \label{fig:device_comparison}
    \vspace{-0.4cm}
\end{figure}

To understand the influence of input recording quality, we analyzed performance metrics across the Professional-device (Pro) and Non-Professional-device (Non-pro) subsets, revealing a significant performance gap as visualized in Figure~\ref{fig:device_comparison}.

The Pro subset consistently yields superior results across all dimensions. It achieves markedly higher scores in perceptual quality metrics (OVRL and speaker similarity), indicating better audio quality and vocal timbre preservation. Furthermore, it demonstrates stronger content intelligibility with a substantially lower and more stable Word Error Rate (WER) compared to the Non-pro subset. This analysis underscores a critical finding: the quality of the recording device is a fundamental factor that significantly impacts the efficacy of singing voice enhancement.

\subsection{Uncovering Performance Trade-offs via Clustering}
\label{sec:cluster_tradeoffs}

\begin{figure}[hbtp]
    \centering
    \begin{minipage}[c]{0.48\textwidth}
        \centering
        \includegraphics[width=\textwidth]{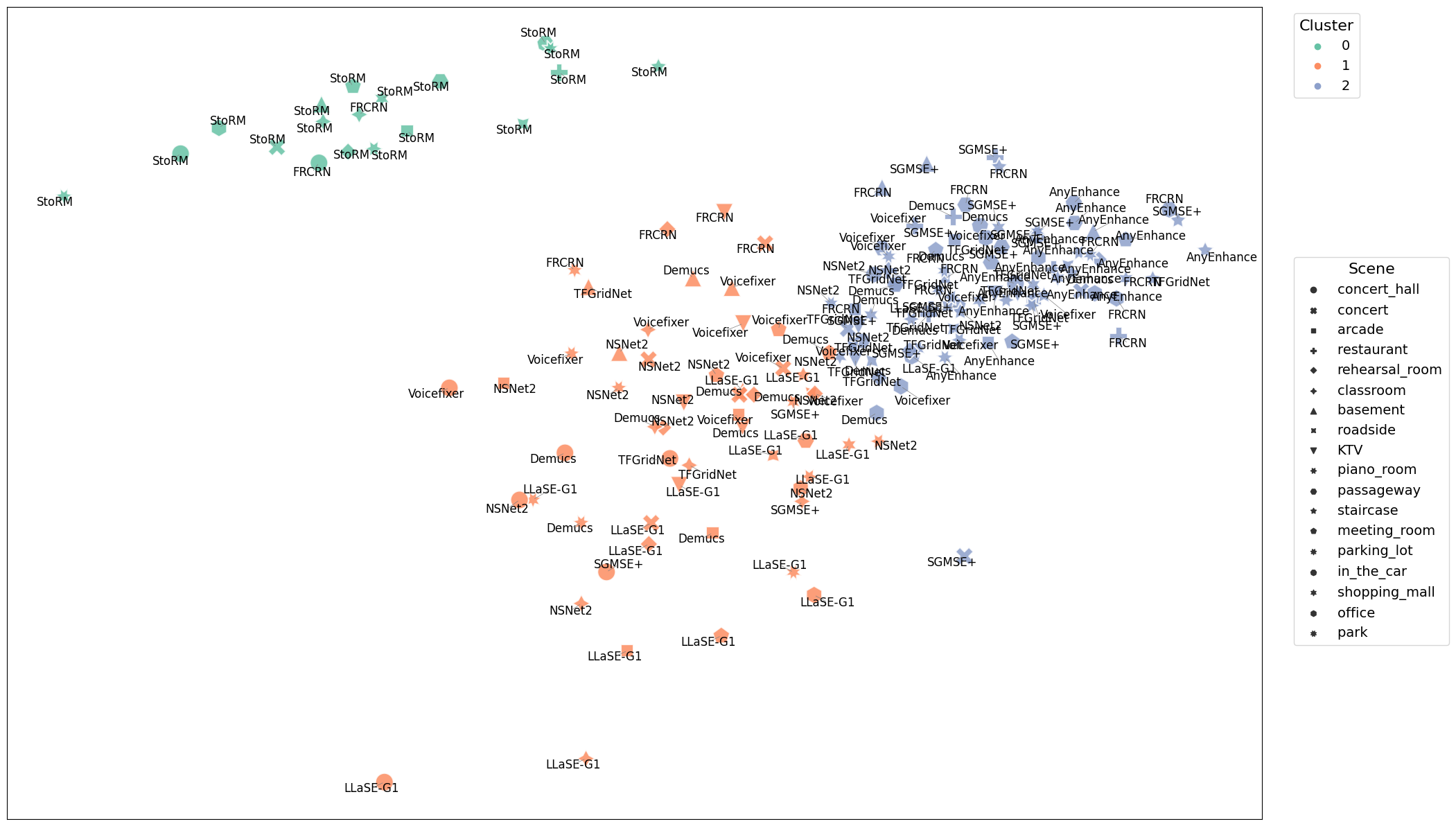}
    \end{minipage}
    \hfill
    \begin{minipage}[c]{0.48\textwidth}
        \centering
        \includegraphics[width=\textwidth]{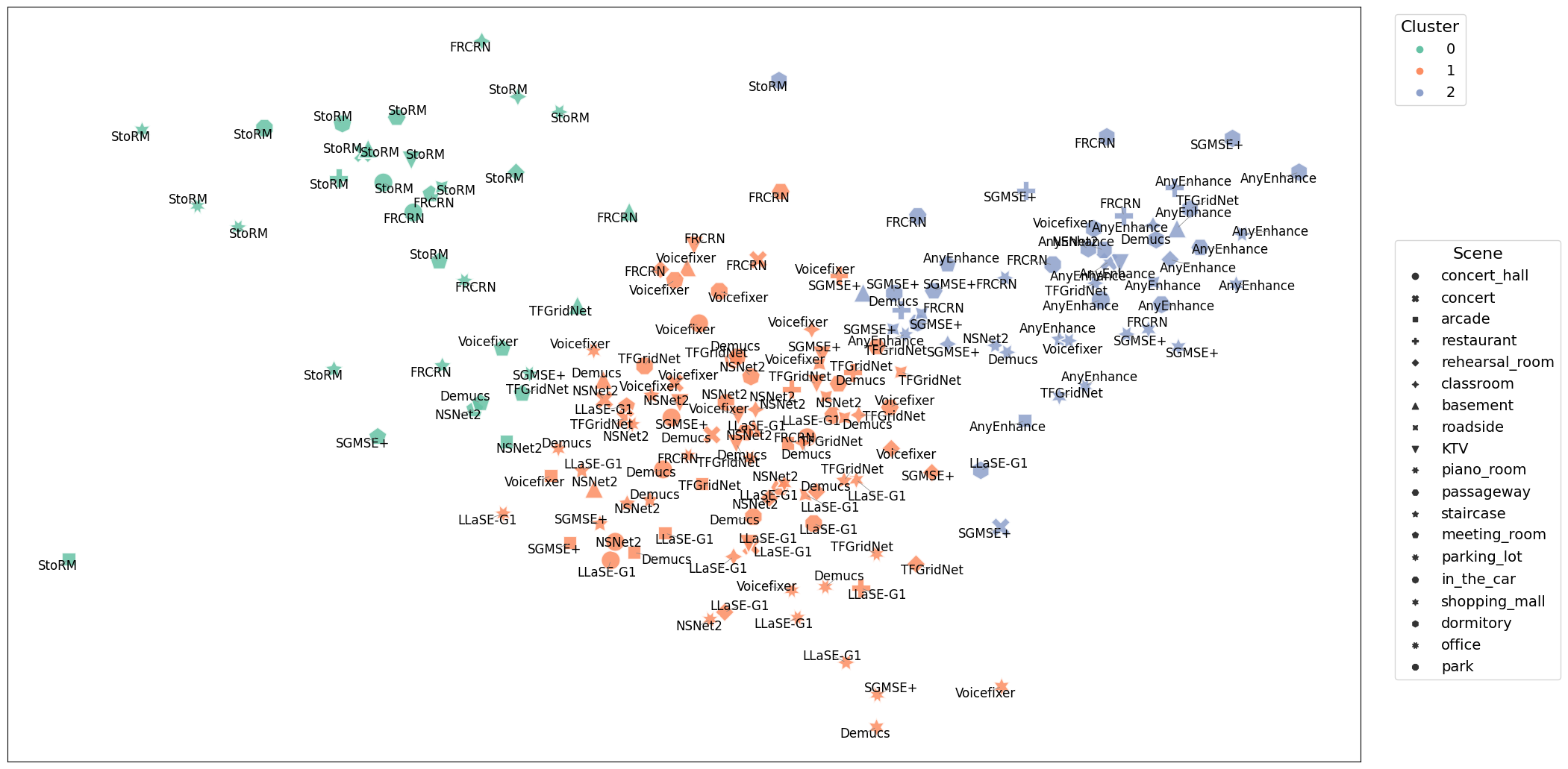}
    \end{minipage}
    \small
    \caption{K-Means clustering of model outputs on the Pro (top) and Non-pro (bottom) subsets, revealing three distinct performance profiles.}
    \label{fig:clustering}
    \vspace{-15pt}
\end{figure}

While Section~\ref{sec:device_impact_analysis} highlights the impact of data quality, a model's own behavior is equally critical. A direct model-by-scenario analysis, however, creates an unwieldy result space (9 models $\times$ 19 scenarios). To manage this complexity and distill core performance patterns, we employ K-Means clustering on the 7-D metric vectors of the enhancement results. After pre-processing the vectors for fair comparison (negating WER and standardizing) and selecting $k=3$ via the elbow method, the analysis revealed three distinct and meaningful clusters, as visualized in Figure~\ref{fig:clustering}.

These clusters, with their centroids detailed in Table~\ref{tab:cluster_means}, expose a prevalent \textbf{trade-off between perceptual quality and content intelligibility}.
\begin{itemize}[itemsep=1pt,topsep=0pt,parsep=0pt]
    \item \textbf{Cluster 0 (Conservative):} Characterized by poor quality but good intelligibility. These models minimally process the audio, preserving content at the cost of effective enhancement.
    \item \textbf{Cluster 1 (Aggressive):} Characterized by good quality but poor intelligibility. These models over-process the audio, achieving higher perceptual scores but severely degrading content.
    \item \textbf{Cluster 2 (Balanced):} Represents the ideal profile, successfully balancing high perceptual quality with excellent content preservation.
\end{itemize}

This finding cautions against relying on single metrics for evaluation. For instance, the favorable WER of Cluster 0 is misleading, as it often reflects inaction rather than successful noise and artifact removal. A model's true effectiveness is therefore better measured by its consistency in achieving the ideal, balanced profile of Cluster 2.

\subsection{Analysis of Model Strategies}
\label{sec:model_strategies}

Our clustering analysis reveals three distinct and consistent strategies adopted by different models. \textbf{Cluster 0} represents a ``conservative'' strategy, characterized by poor perceptual quality but good intelligibility. \storm\ exemplifies this pattern with 100\% of its outputs in this cluster, as it minimally processes the audio and thus fails to provide effective enhancement. In contrast, \textbf{Cluster 1} corresponds to an ``aggressive'' strategy, marked by higher perceptual scores but poor intelligibility. Models such as \llase\ (100\%) and \nsnet\ (89.5\%) predominantly fall here, over-processing the audio to the point of degrading content despite apparent quality gains. \anyenhance, by contrast, achieves an optimal profile (\textbf{Cluster 2}) in 100\% of its outputs, a robustness we attribute to 1) its generative architecture and 2) a training paradigm tailored for general voice enhancement including singing, which addresses the limitations of standard SE models.

\begin{table}[htbp]
    \centering
    \small
    \caption{Mean metric scores for each of the three performance clusters, demonstrating the trade-off between quality (e.g., SIG, OVRL) and intelligibility (WER).}
    \label{tab:cluster_means}
    \renewcommand{\arraystretch}{0.9}
    \setlength{\tabcolsep}{1pt}
    \begin{tabular}{llccccccc}
        \toprule
        \textbf{Device} & \textbf{Cluster} & 
        \textbf{SIG} & \textbf{BAK} & \textbf{OVRL} & 
        \textbf{NISQA} & \textbf{SBERT} & 
        \textbf{SIM} & \textbf{WER} \\
        \midrule
        \multirow{3}{*}{Pro.}
        & 0 & 1.874 & 1.505 & 1.425 & 1.416 & 0.573 & 0.781 & 0.677\\
        & 1 & 2.610 & 3.430 & 2.258 & 2.334 & 0.533 & 0.837 & 2.702\\
        & 2 & 3.144 & 3.627 & 2.749 & 3.225 & 0.653 & 0.890 & 0.897\\        
        \midrule
        \multirow{3}{*}{Non-pro.}
        & 0 & 1.735 & 1.815 & 1.416 & 1.396 & 0.491 & 0.686 & 1.993\\
        & 1 & 2.436 & 3.359 & 2.103 & 2.136 & 0.496 & 0.793 & 3.069\\
        & 2 & 3.044 & 3.523 & 2.635 & 3.162 & 0.618 & 0.857 & 1.307\\
            \midrule
    \end{tabular}
    \vspace{-20pt}
\end{table}

\begin{table}[htbp]
    \centering
    \small
    \caption{Effect of boosting SE models with in-domain singing data. Training with singing voice (`+ Singing`) substantially improves performance on singing tasks (CCMusic GSR, Benchmark) while preserving strong performance on the speech task (LibriVox GSR).}
    \label{tab:result-ablation-singing}
    \setlength{\tabcolsep}{3pt}
    \renewcommand{\arraystretch}{0.9}
    \begin{tabular}{l ccccc}
        \toprule
        \textbf{Model} & 
        \textbf{OVRL}& 
        \textbf{NISQA}& \textbf{SBERT}& 
        \textbf{SIM}& \textbf{WER}\\
        \midrule
        
        \multicolumn{6}{c}{\textbf{CCMusic GSR}} \\
        \midrule
        \voicefixer & 2.501 & 3.022 & 0.550 & 0.810 & 2.613 \\
        \quad + Singing & \textbf{2.508} & \textbf{3.211} & \textbf{0.706} & \textbf{0.864} & \textbf{1.916} \\
        \cmidrule(lr){1-6}
        \anyenhance & 2.559 & 3.039 & 0.569 & 0.799 & 1.876 \\
        \quad + Singing & \textbf{2.761} & \textbf{3.456} & \textbf{0.764} & \textbf{0.893} & \textbf{0.929}\\
        \midrule
        
        \multicolumn{6}{c}{\textbf{LibriVox GSR}} \\
        \midrule
        \voicefixer & 3.029 & \textbf{3.622} & 0.707 & 0.872 & 0.323 \\
        \quad + Singing & \textbf{3.034} & 3.589 & \textbf{0.713} & \textbf{0.876} & \textbf{0.313} \\
        \cmidrule(lr){1-6}
        \anyenhance & 3.094 & 3.920 & 0.736 & 0.898 & \textbf{0.315} \\
        \quad + Singing & \textbf{3.159} & \textbf{4.010} & \textbf{0.736} & \textbf{0.913} & 0.320 \\
        \midrule

        \multicolumn{6}{c}{\textbf{\benchmark}} \\
        \midrule
        \voicefixer & 2.503 & 2.803 & 0.520 & 0.808 & 2.829 \\
        \quad + Singing & \textbf{2.570} & \textbf{3.039} & \textbf{0.611} & \textbf{0.860} & \textbf{1.849} \\
        \cmidrule(lr){1-6}
        \anyenhance & 2.723 & \textbf{3.181} & 0.552 & 0.815 & 2.550 \\
        \quad + Singing & \textbf{2.767} & 3.170 & \textbf{0.643} & \textbf{0.842} & \textbf{1.283} \\
        \bottomrule
    \end{tabular}
    \vspace{-20pt}
\end{table}

\subsection{Boosting Performance with In-Domain Training}
\label{sec:in_domain_training}

Building on our findings—the domain gap (Section.~\ref{sec:motivation}) and the performance analysis (Section~\ref{sec:cluster_tradeoffs})—we hypothesize that in-domain training can significantly improve singing voice enhancement without harming speech performance. We validated this by retraining representative models (\voicefixer and \anyenhance) on a mix of speech and singing data and comparing them to their speech-only counterparts. The results in Table~\ref{tab:result-ablation-singing} confirm our hypothesis: models trained with singing data show marked improvements on singing-related tasks (CCMusic GSR, \benchmark) while their scores on the speech task (LibriVox GSR) remain undiminished. This presents a simple yet effective path for adapting existing SE models to the challenging domain of singing voice.

\section{Conclusion}
\label{sec:conclusion}


In this work, we first quantified the speech-singing domain gap and introduced \textbf{\benchmark}, the first real-world benchmark for this task. Our analysis on \benchmark\ reveals the fundamental impact of device quality and a prevalent quality-intelligibility trade-off that defines distinct model behaviors (e.g., conservative, aggressive). We then validate in-domain training as an effective path to boost singing enhancement without harming speech capabilities. Future work will include expanding \benchmark\ to cover more languages and genres, exploring more perceptually oriented evaluation metrics, and investigating advanced techniques like post-training~\cite{zhang2025multi} to further refine model performance.
\bibliographystyle{IEEEbib}
\bibliography{ref}

\begin{thebibliography}{10}

\bibitem{fabbro2024sound}
Giorgio Fabbro, Stefan Uhlich, Chieh-Hsin Lai, Woosung Choi, Marco
  Mart{\'\i}nez-Ram{\'\i}rez, Weihsiang Liao, Igor Gadelha, Geraldo Ramos,
  Eddie Hsu, Hugo Rodrigues, et~al.,
\newblock ``The sound demixing challenge 2023--music demixing track,''
\newblock {\em TISMIR}, 2024.

\bibitem{li2024robust}
Ruiqi Li, Yu~Zhang, Yongqi Wang, Zhiqing Hong, Rongjie Huang, and Zhou Zhao,
\newblock ``Robust singing voice transcription serves synthesis,''
\newblock in {\em Proc. ACL}, 2024.

\bibitem{zhang2025anyenhance}
Junan Zhang, Jing Yang, Zihao Fang, Yuancheng Wang, Zehua Zhang, Zhuo Wang, Fan
  Fan, and Zhizheng Wu,
\newblock ``{Anyenhance}: A unified generative model with prompt-guidance and
  self-critic for voice enhancement,''
\newblock {\em IEEE TASLP}, 2025.

\bibitem{kang2025llase}
Boyi Kang, Xinfa Zhu, Zihan Zhang, Zhen Ye, Mingshuai Liu, Ziqian Wang, Yike
  Zhu, Guobin Ma, Jun Chen, Longshuai Xiao, et~al.,
\newblock ``{LLaSE-G1}: Incentivizing generalization capability for llama-based
  speech enhancement,''
\newblock {\em arXiv}, 2025.

\bibitem{wang2025metis}
Yuancheng Wang, Jiachen Zheng, Junan Zhang, Xueyao Zhang, Huan Liao, and
  Zhizheng Wu,
\newblock ``Metis: A foundation speech generation model with masked generative
  pre-training,''
\newblock {\em arXiv}, 2025.

\bibitem{xu2023mbtfnet}
Weiming Xu, Zhouxuan Chen, Zhili Tan, Shubo Lv, Runduo Han, Wenjiang Zhou,
  Weifeng Zhao, and Lei Xie,
\newblock ``{MBTFNET}: multi-band temporal-frequency neural network for singing
  voice enhancement,''
\newblock in {\em Proc. ASRU}, 2023.

\bibitem{richter2024ears}
Julius Richter, Yi-Chiao Wu, Steven Krenn, Simon Welker, Bunlong Lay, Shinji
  Watanabe, Alexander Richard, and Timo Gerkmann,
\newblock ``{EARS}: An anechoic fullband speech dataset benchmarked for speech
  enhancement and dereverberation,''
\newblock {\em arXiv}, 2024.

\bibitem{zhang2024gtsinger}
Yu~Zhang, Changhao Pan, Wenxiang Guo, Ruiqi Li, Zhiyuan Zhu, Jialei Wang,
  Wenhao Xu, Jingyu Lu, Zhiqing Hong, Chuxin Wang, et~al.,
\newblock ``Gtsinger: A global multi-technique singing corpus with realistic
  music scores for all singing tasks,''
\newblock in {\em Proc. NeurIPS}, 2024.

\bibitem{liu2022voicefixer}
Haohe Liu, Xubo Liu, Qiuqiang Kong, Qiao Tian, Yan Zhao, DeLiang Wang,
  Chuanzeng Huang, and Yuxuan Wang,
\newblock ``Voicefixer: A unified framework for high-fidelity speech
  restoration,''
\newblock {\em arXiv}, 2022.

\bibitem{wang2023tf}
Zhong-Qiu Wang, Samuele Cornell, Shukjae Choi, Younglo Lee, Byeong-Yeol Kim,
  and Shinji Watanabe,
\newblock ``{TF-GridNet}: Integrating full-and sub-band modeling for speech
  separation,''
\newblock {\em IEEE/ACM TASLP}, 2023.

\bibitem{braun2020data}
Sebastian Braun and Ivan Tashev,
\newblock ``Data augmentation and loss normalization for deep noise
  suppression,''
\newblock in {\em Proc. Interspeech}, 2020.

\bibitem{zhao2022frcrn}
Shengkui Zhao, Bin Ma, Karn~N Watcharasupat, and Woon-Seng Gan,
\newblock ``{FRCRN}: Boosting feature representation using frequency recurrence
  for monaural speech enhancement,''
\newblock in {\em Proc. ICASSP}, 2022.

\bibitem{defossez2019demucs}
Alexandre D{\'e}fossez, Gabriel Synnaeve, and Yossi Adi,
\newblock ``Real time speech enhancement in the waveform domain,''
\newblock in {\em Proc. Interspeech}, 2020.

\bibitem{richter2023sgmse}
Julius Richter, Simon Welker, Jean-Marie Lemercier, Bunlong Lay, and Timo
  Gerkmann,
\newblock ``Speech enhancement and dereverberation with diffusion-based
  generative models,''
\newblock {\em IEEE/ACM TASLP}, 2023.

\bibitem{lemercier2023storm}
Jean-Marie Lemercier, Julius Richter, Simon Welker, and Timo Gerkmann,
\newblock ``Storm: A diffusion-based stochastic regeneration model for speech
  enhancement and dereverberation,''
\newblock {\em IEEE/ACM TASLP}, 2023.

\bibitem{saeki2024speechbertscore}
Takaaki Saeki, Soumi Maiti, Shinnosuke Takamichi, Shinji Watanabe, and Hiroshi
  Saruwatari,
\newblock ``{SpeechBERTScore}: Reference-aware automatic evaluation of speech
  generation leveraging {NLP} evaluation metrics,''
\newblock {\em arXiv}, 2024.

\bibitem{gu2025singnet}
Yicheng Gu, Chaoren Wang, Junan Zhang, Xueyao Zhang, Zihao Fang, Haorui He, and
  Zhizheng Wu,
\newblock ``Singnet: Towards a large-scale, diverse, and in-the-wild singing
  voice dataset,''
\newblock {\em arXiv}, 2025.

\bibitem{reddy2022dnsmos}
Chandan~KA Reddy, Vishak Gopal, and Ross Cutler,
\newblock ``{DNSMOS P.835}: A non-intrusive perceptual objective speech quality
  metric to evaluate noise suppressors,''
\newblock in {\em Proc. ICASSP}, 2022.

\bibitem{mittag2021nisqa}
Gabriel Mittag, Babak Naderi, Assmaa Chehadi, and Sebastian M{\"{o}}ller,
\newblock ``{NISQA}: A deep {CNN}-self-attention model for multidimensional
  speech quality prediction with crowdsourced datasets,''
\newblock in {\em Proc. Interspeech}, 2021.

\bibitem{zhang2025multi}
Junan Zhang, Xueyao Zhang, Jing Yang, Yuancheng Wang, Fan Fan, and Zhizheng Wu,
\newblock ``Multi-metric preference alignment for generative speech
  restoration,''
\newblock {\em arXiv}, 2025.

\end{thebibliography}

\end{document}